\documentclass[aps,prl,twocolumn,superscriptaddress,floatfix,nofootinbib,longbibliography]{revtex4-2}

\usepackage{graphicx}
\usepackage{dcolumn}
\usepackage{bm}
\usepackage{bbm}
\usepackage{amsmath}
\usepackage{amsfonts}
\usepackage{amssymb}
\usepackage{physics}
\usepackage{float}
\usepackage{multirow}
\usepackage{booktabs}
\usepackage{orcidlink}
\usepackage{hyperref}
\hypersetup{
    colorlinks = true,
    citecolor  = blue,
    linkcolor  = blue,
    urlcolor   = blue
}

\begin{document}

\title{\textit{Ab initio} calculations of two-neutrino and neutrinoless double-$\boldsymbol{\beta}$ decay of $^{48}$Ca \\ and related Gamow-Teller strength distributions}

\author{Zhen~Li\,\orcidlink{0000-0003-2786-7272}}
\email{zhen.li1@tu-darmstadt.de}
\affiliation{Technische Universit\"at Darmstadt, Department of Physics, D-64289 Darmstadt, Germany}%
\affiliation{ExtreMe Matter Institute EMMI, GSI Helmholtzzentrum f\"ur Schwerionenforschung GmbH, D-64291 Darmstadt, Germany}
\affiliation{Max-Planck-Institut für Kernphysik, Saupfercheckweg 1, D-69117 Heidelberg, Germany}

\author{Lotta~Jokiniemi\,\orcidlink{0000-0002-9327-5868}}
\email{lotta.jokiniemi@tu-darmstadt.de}
\affiliation{Technische Universit\"at Darmstadt, Department of Physics, D-64289 Darmstadt, Germany}
\affiliation{ExtreMe Matter Institute EMMI, GSI Helmholtzzentrum f\"ur Schwerionenforschung GmbH, D-64291 Darmstadt, Germany}

\author{Achim~Schwenk\,\orcidlink{0000-0001-8027-4076}}
\email{schwenk@physik.tu-darmstadt.de}
\affiliation{Technische Universit\"at Darmstadt, Department of Physics, D-64289 Darmstadt, Germany}
\affiliation{ExtreMe Matter Institute EMMI, GSI Helmholtzzentrum f\"ur Schwerionenforschung GmbH, D-64291 Darmstadt, Germany}
\affiliation{Max-Planck-Institut für Kernphysik, Saupfercheckweg 1, D-69117 Heidelberg, Germany}

\begin{abstract}
We present \textit{ab initio} calculations of two-neutrino double-beta ($2\nu\beta\beta$) decay of $^{48}$Ca and the related Gamow-Teller (GT) strength functions in $^{48}$Sc using the valence-space in-medium similarity renormalization group (VS-IMSRG) with nuclear interactions and electroweak currents based on chiral effective field theory. We find that the usual $pf$-shell valence space significantly underestimates the nuclear matrix element (NME) of $2\nu\beta\beta$ decay compared to experiment, while an enlarged $d_{3/2}pf$ valence space yields very good agreement with the experimental value without any adjustments. We trace this to an improved description of the involved GT strength distributions, so that the enlarged valence space captures important correlations. The enlarged $d_{3/2}pf$ valence space leads to neutrinoless $\beta\beta$ NMEs of $^{48}$Ca that are twice as large compared to the $pf$-shell calculation. Our findings suggest that studies with different valence spaces and related GT strengths are important for assessing \textit{ab initio} NME calculations of heavier $\beta\beta$ decays.
\end{abstract}

\maketitle

\textit{Introduction}---Some even-even nuclei, for which $\beta$ decay is either energetically forbidden or hindered by a large angular-momentum difference---like in the case of $^{48}$Ca~\cite{Kakkar2026}---can undergo double-$\beta$ decay. Double-$\beta$ decay with the emission of two neutrinos ($2\nu\beta\beta$) was first proposed by Maria G\"oppert-Mayer in 1935~\cite{Goeppert-Meyer1935} and has thus far been observed in some dozen nuclei with half-lives ranging between $T_{1/2}^{2\nu}\sim 10^{18}-10^{24}$ years~\cite{Barabash2020,Pritychenko2025,CUORE2025,PandaX2026}, making it the rarest measured nuclear decay.
Neutrinoless double-$\beta$ ($0\nu\beta\beta$) decay is a hypothetical version of the process in which the $\beta$ decays occur without emission of neutrinos. $0\nu\beta\beta$ decay violates lepton-number conservation (without violating baryon number) and requires that neutrinos are Majorana fermions~\cite{Hirsch2006}. Therefore, its observation would have implications on the origin of the matter-antimatter asymmetry of the Universe~\cite{Dolinski2019,Agostini2023,Gomez2023}.

While experiments searching for $0\nu\beta\beta$ decay have reached sensitivities of the order of $T_{1/2}^{0\nu}\sim 10^{26}$ years~\cite{Exo2019,GERDA2020,SNO+2021,CUORE2024,KLZ2025,LEGEND2025} and next-generation experiments aim to reach sensitivities up to $T_{1/2}^{0\nu}\sim 10^{28}$ years~\cite{SNO+2021,nEXO2022,LEGEND2021,CUPID2024,XLZD:2024pdv}, the involved nuclear matrix elements (NMEs) needed for the interpretation of the signals still carry large uncertainties~\cite{Agostini2023}. Most of the predictions are based on various phenomenological methods~\cite{Engel2017,Agostini2023} for which it is difficult to assign reliable uncertainties. With advances in \textit{ab initio} nuclear theory~\cite{Hergert:2020bxy,Hebeler2021,Ekstrom2023}, which solves the many-body problem with systematically improvable approximations based on chiral effective field theory ($\chi$EFT)~\cite{Epelbaum:2008ga,Machleidt:2011zz,Hammer:2019poc,Epelbaum:2019kcf,Cirigliano2018,Cirigliano2018letter,cirigliano2018master}, the $0\nu\beta\beta$-decay NMEs for key candidates have been recently calculated with \textit{ab initio} methods as well~\cite{Yao2020,Belley2021,Novario2021,Belley2023,Belley2024,Lian2026}. Moreover, \textit{ab initio} calculations allow for consistent inclusion of two-body currents (2BCs), which are crucial to reproduce measured single-$\beta$ decays~\cite{Gysbers2019,King:2020wmp,King:2024zbv,Li2026}. Although the effects of these higher-order currents have been estimated in more phenomenological frameworks~\cite{Menendez2011,Wang2018,Jokiniemi2023}, they are yet to be included in \textit{ab initio} calculations of $0\nu\beta\beta$ NMEs~\cite{Chambers-Wall2026}.

The evaluation of $2\nu\beta\beta$-decay NMEs still poses a challenge for \textit{ab initio} calculations, and few results have been reported~\cite{Payne2018,Novario2021,Belley2021,Lian2026}. While $0\nu\beta\beta$-decay operators, which probe higher momentum transfer $q\sim \mathcal{O}(100\,\text{MeV})$, can be evaluated in the so-called closure approximation~\cite{Senkov2014}, the momentum transfer in $2\nu\beta\beta$ decay is restricted by the $Q$-value, $Q_{\beta\beta}\sim\mathcal{O}(1\,\text{MeV})$, making it sensitive to virtual transitions through the intermediate nucleus. These transitions involve excited states up to higher energies and can be difficult to describe~\cite{Payne2018}. However, recent studies have found correlations between the two $\beta\beta$-decay NMEs~\cite{Jokiniemi2023,Horoi2022,Horoi2023,Neacsu2024,Lian2026,Horoi2026}, calling for better understanding of the uncertainties in $2\nu\beta\beta$ calculations.

\begin{figure}[t!]
\centering
\includegraphics[width=\linewidth]{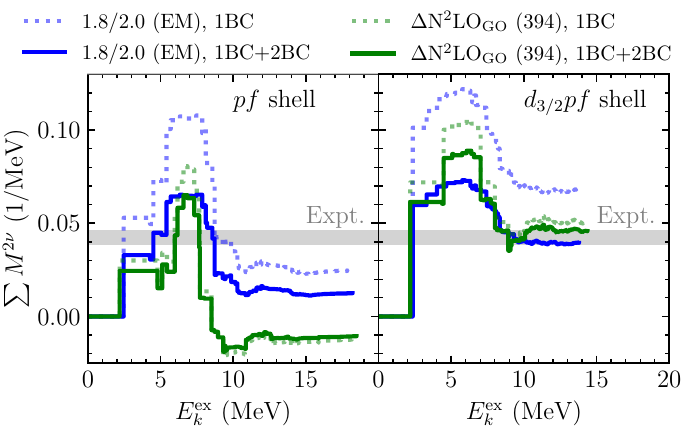}
\caption{Running sum of the $2\nu\beta\beta$ NME for $^{48}$Ca as a function of the excitation energy $E_k^{\rm ex}$ of the intermediate states based on VS-IMSRG(2) calculations with a $pf$ (left panel) or $d_{3/2}pf$ (right panel) valence space. Results are shown with 1BC only (dotted lines) and including 2BC (solid lines) for the 1.8/2.0~(EM)~\cite{Hebeler2011_EM1.8_2.0} and $\Delta$N$^2$LO$_{\rm GO}\,(394)$~\cite{Jiang2020} interactions. The gray band shows the NME extracted from experiment~\cite{Barabash2020}.}
\label{fig:M2nu_running}
\end{figure}

In this Letter, we perform \textit{ab initio} calculations of double-$\beta$ decays of $^{48}$Ca, the lightest $\beta\beta$ emitter, using the valence-space in-medium similarity renormalization group (VS-IMSRG)~\cite{Tsukiyama2011,Tsukiyama2012,Hergert2016,Stroberg2017,Stroberg2019vs} starting from nuclear forces and weak currents derived from $\chi$EFT. We employ two different $\chi$EFT interactions and include the leading 2BCs for $2\nu\beta\beta$ decay. Notably, our results underestimate the NME extracted from the measured $2\nu\beta\beta$ half-life when we perform the calculations in the usual $pf$-shell valence space, but an enlarged $d_{3/2}pf$ valence space produces results consistent with experiment for both $\chi$EFT interactions without adjustments, see Fig.~\ref{fig:M2nu_running}. This can be traced back to an improved description of the involved Gamow-Teller (GT) strength distributions. Finally, we calculate the $0\nu\beta\beta$ NME in the enlarged valence space and show that the NME is increased to twice the value in the $pf$ shell, thus leading to a four-times shorter $0\nu\beta\beta$ half-life.

\textit{Theoretical framework}---The $2\nu\beta\beta$ half-life at the leading order is given by $\big(T^{2\nu}_{1/2}\big)^{-1} =  G^{2\nu} g_\mathrm{A}^4 \bigl|M^{2\nu} m_e\bigr|^2$~\cite{Suhonen1998,Engel2017,elMorabit2025}, with phase-space factor $G^{2\nu}$~\cite{Kotila2012,Kotila2013}, axial-vector coupling $g_{\rm A}$, and electron mass $m_e$. The $2\nu\beta\beta$ NME is
\begin{align}\label{eq:MGT2nu}
M^{2\nu} 
= \sum_{k} \frac{\langle 0_{\rm f}^{+}||\mathbf{GT}^-||1_k^+\rangle 
\langle 1_k^+ || \mathbf{GT}^- || 0_{\rm i}^+ \rangle}
{E_{k} - (E_{\rm gs}^{\rm i} + E_{\rm gs}^{\rm f})/2} \,, 
\end{align}
where $|0_{\rm i}^+\rangle$ and $|0_{\rm f}^+\rangle$ are the ground states of the initial (i) and final (f) nuclei with the respective energies $E_{\rm gs}^{\rm i}$ and $E_{\rm gs}^{\rm f}$, and $|1_k^+\rangle$ is the $k$th intermediate $1^+$ state with energy $E_k$ (the corresponding excitation energy is denoted as $E_k^{\rm ex}$). 
The GT operator is $\mathbf{GT}^- = ({\bf J}_x + i {\bf J}_y) / g_{\rm A}$ with isospin components ${\bf J}_x$ and ${\bf J}_y$ of the spatial axial-vector current ${\bf J}$ at vanishing momentum transfer. 
In addition to the one-body current (1BC), we include the leading 2BCs at next-to-next-to leading order (N$^2$LO) from $\chi$EFT~\cite{Park2003,Krebs2017,Hoferichter2020}, 
following Ref.~\cite{Li2026}.  
We do not consider other higher-order corrections whose effects are expected to remain below 10\%~\cite{Simkovic2018b,elMorabit2025,Castillo2026}. We also neglect the Fermi contribution to the NME~\cite{Suhonen1998}, since the double Fermi transition mainly occurs between isobaric analog states and is therefore suppressed for the ground-state-to-ground-state transition for $^{48}$Ca.

Our results start from two different $\chi$EFT interactions, 1.8/2.0~(EM)~\cite{Hebeler2011_EM1.8_2.0} and $\Delta$N$^2$LO$_{\rm GO}\,(394)$~\cite{Jiang2020}, which include nucleon-nucleon (NN) and three-nucleon (3N) contributions, whose low-energy constants (LECs) are fitted to NN scattering and few-body observables, and informed by properties of heavier nuclei and nuclear matter in the case of $\Delta$N$^2$LO$_{\rm GO}\,(394)$. The same 3N LECs and regulator scale are used for the 2BCs~\cite{Li2026}. The $\Delta$N$^2$LO$_{\rm GO}\,(394)$ interaction includes $\Delta$ excitations explicitly, as is the case for the corresponding leading 2BCs. 

The VS-IMSRG~\cite{Tsukiyama2011,Tsukiyama2012,Hergert2016,Stroberg2017,Stroberg2019vs} is used to compute the nuclear states and energies for the NME calculation involved in Eq.~\eqref{eq:MGT2nu}, without any phenomenological adjustments. The VS-IMSRG performs a series of unitary transformations of the many-body Hamiltonian $H$ to decouple the correlations between a selected valence space and its complement. Hence, the eigenvalue problem of the Hamiltonian in the full model space is transformed to a tractable eigenvalue problem of an effective Hamiltonian (i.e., the evolved Hamiltonian) in the selected valence space, which can be then diagonalized with standard shell-model diagonalization methods. The particle-hole excitations from the valence space to its complement are introduced in the effective Hamiltonian through the VS-IMSRG unitary transformations. During the unitary transformation, the Hamiltonian and all operators are truncated to the normal-ordered two-body level [referred to as VS-IMSRG(2)]~\cite{Stroberg2017}. The calculation is built from a Hartree-Fock basis of 13 major harmonic oscillator shells, i.e., $e_{\rm max}={\rm max}(2n+l)=12$, with frequency $\hbar\omega=16$~MeV, unless otherwise stated. The 3N matrix elements are further restricted by $E_{\rm 3\,max}=24$, which is sufficiently large for the nuclei with $A=48$~\cite{Miyagi2022_E3max,Belley2021}.  
The GT operator is transformed accordingly using the initial nucleus as a reference for the normal ordering. 
The Hamiltonian and 2BCs are generated with the \texttt{NuHamil}~\cite{Miyagi2023NuHamil} code, and the VS-IMSRG calculation is carried out with the codes \texttt{IMSRG++}~\cite{imsrgpp} and \texttt{KSHELL}~\cite{Shimizu2019}. 

\textit{Results for $2\nu\beta\beta$ decay}---We first consider the $pf$ shell as the valence space on top of a $^{40}$Ca core. This has also been used for VS-IMSRG calculations of $^{48}$Ca $\beta\beta$ decays~\cite{Payne2018,Belley2021}. The running sum of the $2\nu\beta\beta$ NME $M^{2\nu}$ is shown in the left panel of Fig.~\ref{fig:M2nu_running}, where up to 250 intermediate $1^+$ states are included. The NMEs given by both the 1.8/2.0~(EM) and $\Delta$N$^2$LO$_{\rm GO}$\,(394) interactions are significantly smaller than the value extracted from experiment~\cite{Barabash2020} especially with the inclusion of 2BCs for the 1.8/2.0~(EM) interaction.
The small NME from VS-IMSRG calculations was also reported in Refs.~\cite{Payne2018,Belley2021} for the 1.8/2.0~(EM) interaction.
A larger reduction of the matrix elements due to 2BCs is observed with the 1.8/2.0~(EM) interaction than with the $\Delta$N$^2$LO$_{\rm GO}$\,(394) interaction, yielding similar NMEs in magnitude, although with different signs, from these two interactions. The running sum increases up to $E_k^{\rm ex} \approx 7$\,MeV and then decreases up to $E_k^{\rm ex} \approx 10$\,MeV. The decrease is much more pronounced for the $\Delta$N$^2$LO$_{\rm GO}$\,(394) interaction than for the 1.8/2.0~(EM) interaction so that the NME for the former case is eventually negative.

The right panel of Fig.~\ref{fig:M2nu_running} shows our results from the same calculation but with the $d_{3/2}pf$ shell as valence space (i.e., the $0d_{3/2}$ orbital plus the $pf$ shell on top of a $^{32}$S core). 
For this multishell valence space, the spurious center-of-mass (c.m.) motion is removed by modifying the Hamiltonian as $H'=H+\beta H_{\rm c.m.}$ with c.m. Hamiltonian $H_{\rm c.m.}$ and parameter $\beta=2$~\cite{Gloeckner1974,Miyagi2020}.
In addition, an energy shift $\Delta=5$~MeV is introduced in the denominator of the IMSRG generator~\cite{Miyagi2020}.
To make the calculation feasible, a $4\hbar\omega$ truncation (i.e., limiting the number of nucleons excited from the $0d_{3/2}$ orbital to the $pf$ shell to 4) is employed during the diagonalization of the valence-space Hamiltonian, and 250 intermediate states are included in the running sum.  
Notably, with this enlarged valence space, the increase of the running sum becomes more pronounced, and the decrease is reduced. Remarkably, we find a very good agreement with the value extracted from experiment~\cite{Barabash2020} for both the 1.8/2.0~(EM) and $\Delta$N$^2$LO$_{\rm GO}\,(394)$ interactions when 2BCs are included. A large discrepancy is observed between the two interactions when only 1BCs are included.

\begin{figure}[t!]
\centering
\includegraphics[width=\linewidth]{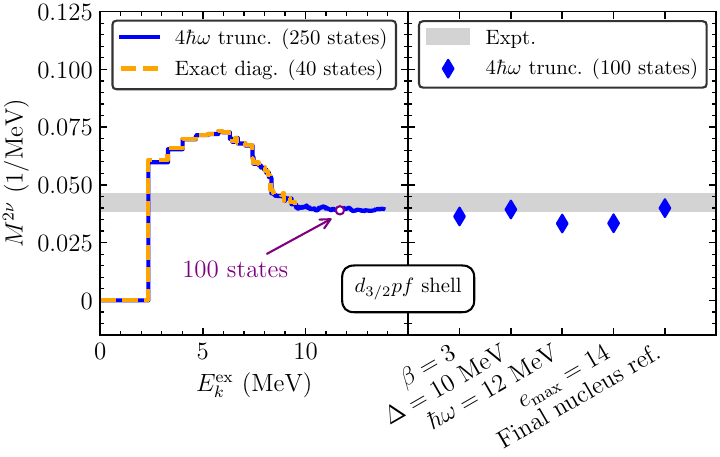}
\caption{$2\nu\beta\beta$-decay NME of $^{48}$Ca from VS-IMSRG(2) calculations with the $d_{3/2}pf$ valence space using the 1.8/2.0~(EM) interaction~\cite{Hebeler2011_EM1.8_2.0} and including 2BCs. Results are shown to test the sensitivity to calculational choices. Left panel: Running sum from the $4\hbar\omega$ truncation with 250 intermediate states (blue solid line) and from exact diagonalization with 40 intermediate states (orange dashed line), using $\beta=2$, $\Delta=5$\,MeV, $\hbar\omega=16$\,MeV, $e_{\rm max}=12$, and initial nucleus reference to normal order and evolve the GT 1BC+2BC operator. Right panel: NMEs obtained by individually varying different parameters, to $\beta=3$, to $\Delta=10$\,MeV, to $\hbar\omega=12$\,MeV, to $e_{\rm max}=14$, and to final nucleus reference, based on the $4\hbar\omega$ truncation and 100 intermediate states.}
\label{fig:M2nu_uncertainty}
\end{figure}

To evaluate the robustness of the improvement with the enlarged $d_{3/2}pf$ valence space, we show in Fig.~\ref{fig:M2nu_uncertainty} results for the 1.8/2.0~(EM) interaction with 2BCs when the different calculational choices are varied.
We observe that the $4\hbar\omega$ truncation provides a very good approximation compared to the exact diagonalization, and no significant changes are observed by varying the other basis and VS-IMSRG parameters, suggesting a solid improvement with the enlarged valence space.

\begin{figure*}[t!]
\centering
\includegraphics[width=0.875\linewidth]{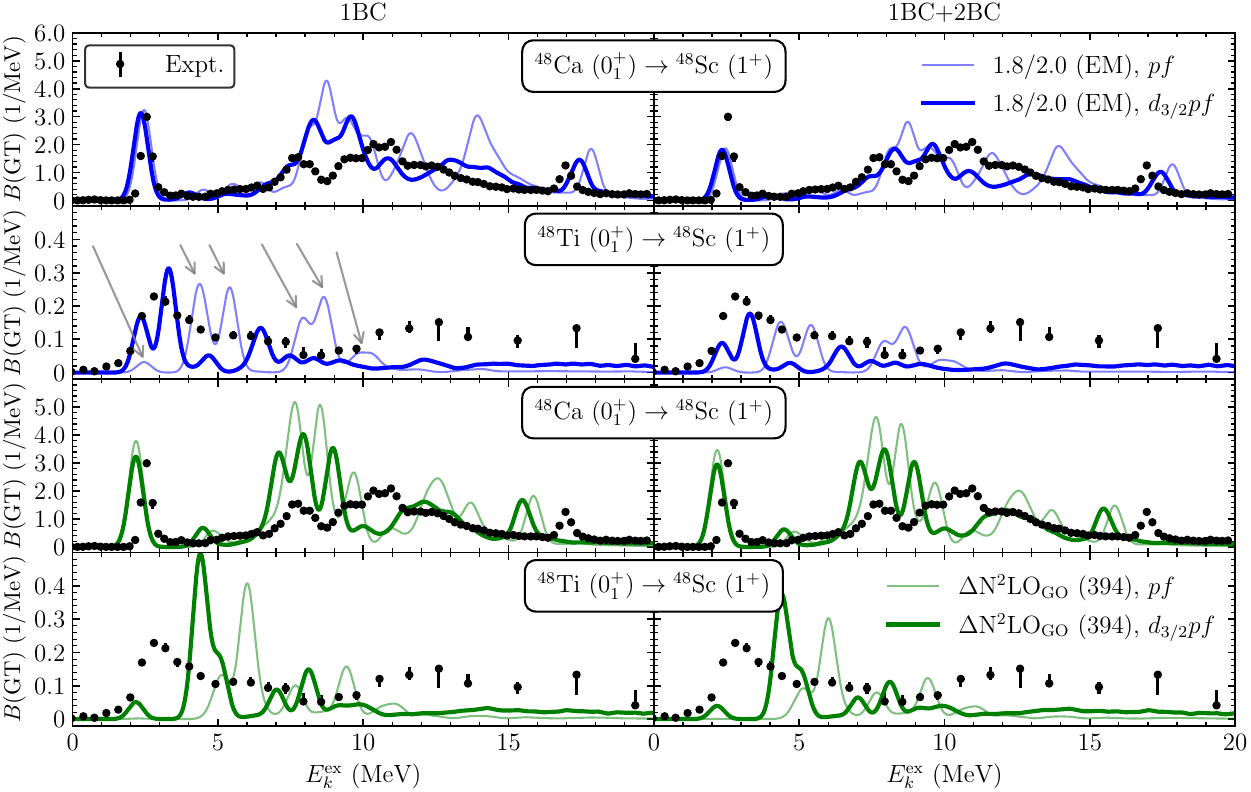}
\caption{VS-IMSRG(2) results for GT transition strength distributions with the $pf$ and $d_{3/2}pf$ valence spaces for ${}^{48}{\rm Ca}~(0_1^+)\rightarrow{}^{48}{\rm Sc}~(1_{k}^+)$ and ${}^{48}{\rm Ti}~(0_1^+)\rightarrow{}^{48}{\rm Sc}~(1_{k}^+)$ using the 1.8/2.0~(EM)~\cite{Hebeler2011_EM1.8_2.0} (upper two rows) and $\Delta$N$^2$LO$_{\rm GO}\,(394)$~\cite{Jiang2020} (lower two rows) interactions with 1BC (left panels) and 1BC+2BC (right panels) as a function of the excitation energy $E_k^{\rm ex}$ of the final $1_{k}^+$ states. The Lanczos strength function method~\cite{Caurier2005,Haxton2005} is used with 200 iterations to calculate the GT strengths, which are smoothed with a Gaussian with $0.25$\,MeV width. Results are compared to GT strength distributions extracted from experimental charge-exchange reactions~\cite{Yako2009}.}
\label{fig:GTs}
\end{figure*}

\textit{GT strengths and charge-exchange reactions}---The improvement due to the enlarged valence space can be understood by inspecting the GT transition probabilities $B({\rm GT}) = |\langle 1_{k}^+ || {\rm GT} || 0_{1}^+ \rangle|^2/(2J+1)$ (with angular momentum of initial state $J=0$ here) involved in the calculation of the $2\nu\beta\beta$ NME in Eq.~(\ref{eq:MGT2nu}), i.e., $^{48}$Ca\,($0_{\rm i}^+$)\,$\rightarrow$\,$^{48}$Sc\,($1_{k}^+$) and $^{48}$Ti\,($0_{\rm f}^+$)\,$\rightarrow$\,$^{48}$Sc\,($1_{k}^+$), and comparing them with experimental charge-exchange reactions. We also include the 2BCs to facilitate comparisons with the running sums, though we note that the 2BCs derived for weak interactions may not be fully appropriate for charge-exchange reactions.

Figure~\ref{fig:GTs} shows the smoothed $B({\rm GT})$ distributions compared to the GT strength distributions extracted from $^{48}{\rm Ca}(p,\,n)$ and $^{48}{\rm Ti}(n,\,p)$ reactions~\cite{Yako2009}. 
The GT transition strength from $^{48}$Ti to $^{48}$Sc computed in the $pf$ valence space with the 1.8/2.0~(EM) interaction and 1BC only has six main peaks, marked by arrows in the figure (second left panel). 
There are also peaks at the corresponding $E_k^{\rm ex}$ in the transition from $^{48}$Ca to $^{48}$Sc, although the strength is more distributed to higher energies. 
These peaks produce six large changes in the running sum shown in Fig.~\ref{fig:M2nu_running}: The increase of the running sum is mainly driven by the first three peaks, while the decrease is caused by the last three peaks. 
The strengths above $E_k^{\rm ex} \approx 10$\,MeV from $^{48}$Ca to $^{48}$Sc do not contribute much to the running sum due to the large energy denominator and the suppression of the corresponding strength from $^{48}$Ti to $^{48}$Sc.   
With the enlarged $d_{3/2}pf$ valence space, the largest peaks in the transition strength from $^{48}$Ti to $^{48}$Sc are shifted to lower energies, while the strength extends to higher energies $E_k^{\rm ex}>10$\,MeV. 
The transition strength from $^{48}$Ca to $^{48}$Sc with the enlarged valence space changes less dramatically for $E_k^{\rm ex} \lesssim 10$\,MeV, however the strength at around 8.5\,MeV is reduced by around 50\%. 
In sum, we find that the GT transition strength is in general enhanced for $E_k^{\rm ex} \lesssim 7$\,MeV while weakened for $7\,\text{MeV}\lesssim E_k^{\rm ex} \lesssim 10$\,MeV with the enlarged valence space, leading to the more pronounced increase and less decrease in the running sum of $M^{2\nu}$. 
This finding also holds when 2BCs are included or when the $\Delta$N$^2$LO$_{\rm GO}\,(394)$ interaction is used, see Fig.~\ref{fig:GTs}. 

The virtue of the enlarged valence space in the calculation of $2\nu\beta\beta$-decay NME can be validated by a comparison to the transition strength extracted from charge-exchange reactions~\cite{Yako2009}. 
As shown in Fig.~\ref{fig:GTs}, the transition strength with the enlarged valence space matches better the experimental strength.
Specifically, the peaks above $7$\,MeV become broader and less pronounced for the transitions from $^{48}$Ca to $^{48}$Sc, leading to improved agreement with experiment for both interactions. 
For the transitions from $^{48}$Ti to $^{48}$Sc, the positions of the peaks move toward the experimental ones, although there is still a large mismatch for the $\Delta$N$^2$LO$_{\rm GO}\,(394)$ interaction. 
In general, we observe that the 1.8/2.0~(EM) interaction yields better agreement with experiment compared to the $\Delta$N$^2$LO$_{\rm GO}\,(394)$ interaction, while both interactions underestimate the transition strength at around $10\,\text{MeV} \lesssim E_k^{\rm ex} \lesssim 12$\,MeV from $^{48}$Ca to $^{48}$Sc and at $E_k^{\rm ex} \gtrsim 10$\,MeV from $^{48}$Ti to $^{48}$Sc. 
It is worth mentioning that the $1^+$-state spectrum in $^{48}$Sc gets compressed with the $d_{3/2}pf$ space, in better agreement with experiment.

\begin{figure*}[t!]
\centering
\includegraphics[width=0.775\linewidth]{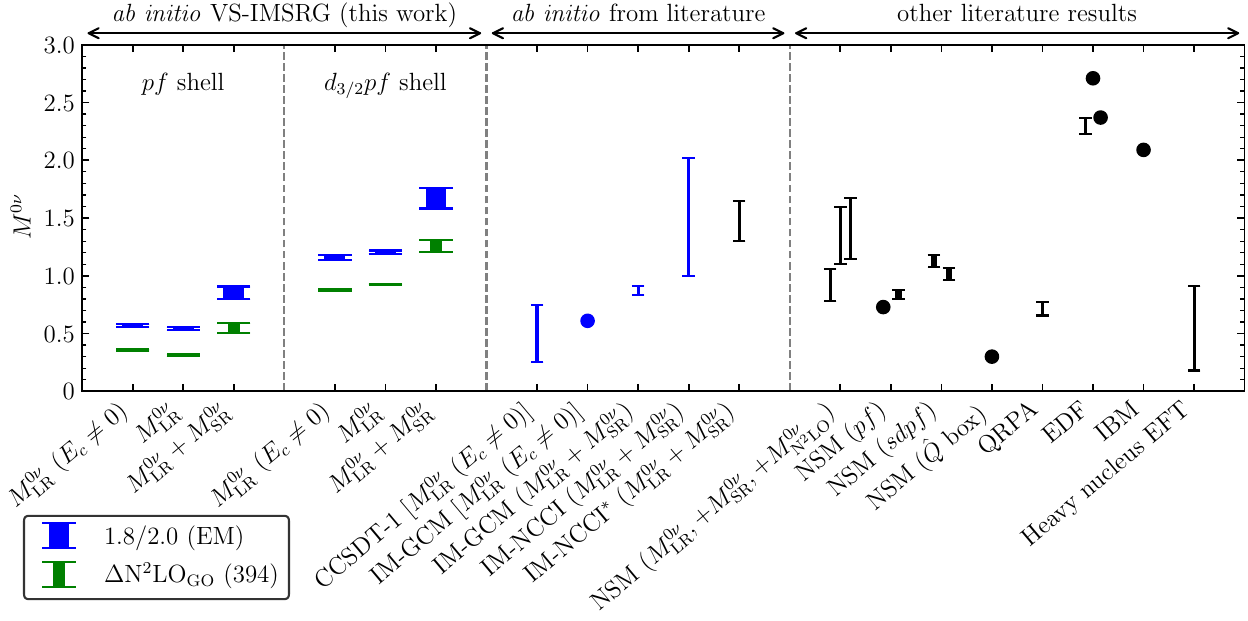}
\caption{$0\nu\beta\beta$-decay NMEs from VS-IMSRG(2) calculations with the $pf$ and $d_{3/2}pf$ valence spaces using the 1.8/2.0~(EM)~\cite{Hebeler2011_EM1.8_2.0} and $\Delta$N$^2$LO$_{\rm GO}\,(394)$~\cite{Jiang2020} interactions (left two panels).
The NME ranges for each interaction are based on $e_{\rm max}=12$ with the initial reference and $e_{\rm max}=14$ with the final and initial references to probe the uncertainty, including also the SR uncertainty~\cite{Wirth2021} for $M_{\rm LR}^{0\nu}+M_{\rm SR}^{0\nu}$. Our results are compared to other \textit{ab initio} calculations (third panel) with the 1.8/2.0~(EM) interaction (blue symbols) based on coupled-cluster theory with singles-doubles-and-leading-triples excitations (CCSDT-1)~\cite{Novario2021}, the IMSRG with the generator coordinate method (IM-GCM)~\cite{Yao2020,Wirth2021}, the in-medium no-core configuration-interaction (IM-NCCI) method~\cite{Lian2026}, as well as with the non-implausible interaction samples~\cite{Hu2022ab} using IM-NCCI conditioned on the experimental value of the $2\nu\beta\beta$ NME (referred to as IM-NCCI$^*$)~\cite{Lian2026}. 
We also compare to the recent nuclear shell model (NSM) calculation of $M_{\rm LR}^{0\nu}$, $M_{\rm LR}^{0\nu}+M_{\rm SR}^{0\nu}$, and $M_{\rm LR}^{0\nu}+M_{\rm SR}^{0\nu}+M_{\rm N^2LO}^{0\nu}$ (from left to right)~\cite{Castillo2025}, as well as other calculations of $M^{0\nu}_{\rm LR}(E_c\neq 0)$ based on the NSM in the $pf$~\cite{Senkov2013,Iwata2016} and $sdpf$ shell~\cite{Iwata2016} (consistent short-range correlations are used for the thicker error bars), the NSM in combination with the $\hat{Q}$ box method~\cite{Coraggio2020}, 
the quasiparticle random phase approximation (QRPA)~\cite{Simkovic2018}, energy density functional (EDF) theory~\cite{Rodriguez2010,LopezVaquero2013,Song2017}, the interacting boson model (IBM)~\cite{Barea2015,Deppisch2020}, and heavy nucleus EFT~\cite{Brase2022}.}
\label{fig:M0nu}
\end{figure*}

\textit{Implications for $0\nu\beta\beta$ decay}---The above results indicate that neglected higher-body operators in the VS-IMSRG(2) calculation with the $pf$-shell valence space carry important correlations for $M^{2\nu}$, while some of these can be introduced with the enlarged $d_{3/2}pf$ valence space. 
It is therefore of great interest to see how the additional correlations captured by the enlarged valence space impact the $0\nu\beta\beta$ NME, also since the $2\nu\beta\beta$ and $0\nu\beta\beta$ NMEs have been found to be correlated~\cite{Jokiniemi2023,Horoi2022,Horoi2023,Neacsu2024,Lian2026,Horoi2026}.

In the light-neutrino-exchange mechanism, the half-life of $0\nu\beta\beta$ decay is given by $\big(T^{0\nu}_{1/2}\big)^{-1} =  G^{0\nu} g_\mathrm{A}^4 \bigl| M^{0\nu} \frac{m_{\beta\beta}}{m_e} \bigr|^2$ \cite{Agostini2023} with effective Majorana mass $m_{\beta\beta}$ and phase-space factor $G^{0\nu}$~\cite{Kotila2012}.
The $0\nu\beta\beta$ NME at leading order in  $\chi$EFT~\cite{Cirigliano2018,cirigliano2018master,Cirigliano2018letter} is written as $M_{}^{0\nu} = M_{\rm LR}^{0\nu} + M_{\rm SR}^{0\nu}$, with a long-range (LR) term $M_{\rm LR}^{0\nu}$ and a short-range (SR) contact term $M_{\rm SR}^{0\nu}$. The LR term (with N$^2$LO corrections to the 1BCs) has the same expression as in the phenomenological formalism~\cite{Engel2017} but without the closure energy $E_c$ in the denominator (or with $E_c=0$). We take the unknown coupling in the SR term from the matching procedure in Ref.~\cite{Wirth2021} with the synthetic data from Refs.~\cite{Cirigliano2021,cirigliano2021b}. 
Here, we exclude higher-order corrections, such as contributions from ultrasoft neutrinos and loops~\cite{Cirigliano2018} (whose effects are negligible for $^{48}$Ca in phenomenological shell model calculations~\cite{Castillo2025}) and from recently derived three-nucleon potentials~\cite{Chambers-Wall2026}. 
To facilitate comparison with previous LR term calculations using $E_c\neq0$ [denoted as $M_{\rm LR}^{0\nu}(E_c\neq0)$], we also carry out calculations with the commonly used $E_c=7.72$\,MeV~\cite{Tomoda1991}.

Our results are shown in Fig.~\ref{fig:M0nu} (see also Table~\ref{tab:M0nu} in End Matter) with a comparison to \textit{ab initio} and other calculations from the literature.
We achieve a full diagonalization of the valence-space Hamiltonian for the calculation of $0\nu\beta\beta$ NME.  
With the $pf$-shell valence space, our calculations match the VS-IMSRG values reported in Refs.~\cite{Payne2018,Belley2021,Belley2024thesis} and are consistent with the results from IM-GCM~\cite{Yao2020,Wirth2021} and CCSDT-1~\cite{Novario2021}. With the enlarged $d_{3/2}pf$ valence space, the LR NME $M_{\rm LR}^{0\nu}$ for the 1.8/2.0 interaction is twice as large, while the result for the $\Delta$N$^2$LO$_{\rm GO}\,(394)$ interaction is three times larger. For both interactions, the combined LR+SR NMEs is two times larger. The value of $M_{\rm LR}^{0\nu}+M_{\rm SR}^{0\nu}$ with the enlarged valence space better matches the IM-NCCI result~\cite{Lian2026}, which includes three-particle--three-hole configurations in $^{48}$Sc shown to be important for these transitions. 

As shown in Fig.~\ref{fig:M0nu}, the results of $M_{\rm LR}^{0\nu}(E_c\neq0)$ are very close to those evaluated with the $\chi$EFT expression. Compared to the $\textit{ab initio}$ LR NMEs from CCSDT-1~\cite{Novario2021} and IM-GCM~\cite{Yao2020} calculations, our extended space results also yield larger LR NMEs, while our $pf$-shell results are similar (the same is true for the SR term when compared to IM-GCM).
Regarding other literature values, our $d_{3/2}pf$ valence space NMEs are generally larger than the nuclear shell model from Refs.~\cite{Senkov2013,Iwata2016,Coraggio2020}, QRPA~\cite{Simkovic2018}, and heavy nucleus EFT results~\cite{Brase2022}, similar in size to the recent NSM results~\cite{Castillo2025}, but smaller than those from 
EDF~\cite{Rodriguez2010,LopezVaquero2013,Song2017} and IBM calculations~\cite{Barea2015,Deppisch2020}. Finally, an increase (albeit smaller) from $pf$ to the extended $sdpf$ space was also found in phenomenological NSM calculations~\cite{Iwata2016} (see the thicker error bars).

In this Letter, we have carried out \textit{ab initio} VS-IMSRG calculations for the $2\nu\beta\beta$- and $0\nu\beta\beta$-decay NMEs of $^{48}$Ca starting from $\chi$EFT NN+3N interactions, including 2BCs up to N$^2$LO for $2\nu\beta\beta$ decay. Our results show that correlations important for the $2\nu\beta\beta$ calculation are missing with the usual $pf$-shell valence space, whereas an enlarged $d_{3/2}pf$ valence space leads to good agreement with the experimental $2\nu\beta\beta$ NME and provides an improved description of the strength functions in $^{48}$Sc extracted from charge-exchange reactions. The enlarged $d_{3/2}pf$ valence space leads to $0\nu\beta\beta$ NMEs of $^{48}$Ca that are approximately twice as large compared to the corresponding $pf$-shell calculation. Our results show that the additional correlations captured by enlarged valence spaces are crucial for $0\nu\beta\beta$ decay. This calls for \textit{ab initio} studies with extended valence spaces~\cite{Parzuchowski2017,Miyagi2020} and related GT strengths for $0\nu\beta\beta$-decay NMEs of the key candidates, such as $^{76}$Ge, $^{100}$Mo, $^{130}$Te, and $^{136}$Xe. Another important direction is the inclusion of higher-body operators in the IMSRG~\cite{Heinz2021,He2024,Stroberg2024,Heinz:2024juw} and three-body contributions stemming from 2BC and other effects~\cite{Chambers-Wall2026}.
 
\begin{acknowledgments}
We thank Martin Hoferichter and Javier Men\'endez for useful discussions. This work was supported in part by the European Research Council (ERC) under the European Union’s Horizon 2020 research and innovation programme (Grant Agreement No.~101020842) and by the LOEWE Top Professorship LOEWE/4a/519/05.00.002(0014)98 by the State of Hesse. The authors gratefully acknowledge the Gauss Centre for Supercomputing e.V. for providing computing time through the John von Neumann Institute for Computing (NIC) on the GCS Supercomputer JUWELS at J\"{u}lich Supercomputing Centre (JSC), as well as computing time provided on the high-performance computer Lichtenberg II at TU Darmstadt.
\end{acknowledgments}

\bibliography{two-nu}

\appendix

\onecolumngrid

\section{End Matter}

In Tables~\ref{tab:M2nu}~and~\ref{tab:M0nu}, we provide the values of NMEs shown in Figs.~\ref{fig:M2nu_running}~and~\ref{fig:M0nu}. 

\begin{table}[h!]
\caption{\label{tab:M2nu}Results for the NMEs $|M^{2\nu}|$ (MeV$^{-1}$) of $2\nu\beta\beta$ decay shown in Fig.~\ref{fig:M2nu_running} for the 1.8/2.0~(EM)~\cite{Hebeler2011_EM1.8_2.0} and $\Delta$N$^2$LO$_{\rm GO}\,(394)$~\cite{Jiang2020} interactions.}
\begin{tabular}{lcccc}
\toprule
\multirow{2}{*}{} & \multicolumn{2}{c}{$pf$ shell} & \multicolumn{2}{c}{$d_{3/2}pf$ shell} \\
\cmidrule(lr){2-3}
\cmidrule(lr){4-5}
& 1.8/2.0~(EM) & $\Delta$N$^2$LO$_{\rm GO}\,(394)$ & 1.8/2.0~(EM) & $\Delta$N$^2$LO$_{\rm GO}\,(394)$ \\
\midrule
1BC     & 0.0246 & 0.0124 & 0.0681 & 0.0504 \\
1BC+2BC & 0.0125 & 0.0106 & 0.0395 & 0.0458 \\
\bottomrule
\end{tabular}
\end{table}

\begin{table}[h!]
\caption{\label{tab:M0nu}Results for the NMEs $M_{\rm LR}^{0\nu}~(E_c\neq0)$, $M_{\rm LR}^{0\nu}$, and $M_{\rm LR}^{0\nu}+M_{\rm SR}^{0\nu}$ of $0\nu\beta\beta$ decay shown in Fig.~\ref{fig:M0nu} for the 1.8/2.0~(EM)~\cite{Hebeler2011_EM1.8_2.0} and $\Delta$N$^2$LO$_{\rm GO}\,(394)$~\cite{Jiang2020} interactions. The NME range is from the calculations using $e_{\rm max}=12$ with the initial nucleus reference as well as $e_{\rm max}=14$ with the initial nucleus and final nucleus references.}
\begin{tabular}{lcccc}
\toprule
\multirow{2}{*}{} & \multicolumn{2}{c}{$pf$ shell} & \multicolumn{2}{c}{$d_{3/2}pf$ shell} \\
\cmidrule(lr){2-3}
\cmidrule(lr){4-5}
& 1.8/2.0~(EM) & $\Delta$N$^2$LO$_{\rm GO}\,(394)$ & 1.8/2.0~(EM) & $\Delta$N$^2$LO$_{\rm GO}\,(394)$ \\
\midrule
$M_{\rm LR}^{0\nu}~(E_c\neq0)$ & 0.557--0.579 & 0.351--0.366 & 1.135--1.180 & 0.872--0.886 \\
$M_{\rm LR}^{0\nu}$ & 0.528--0.556 & 0.302--0.322 & 1.192--1.219 & 0.924--0.929 \\
$M_{\rm LR}^{0\nu}+M_{\rm SR}^{0\nu}$ & 0.796--0.907 & 0.507--0.589 & 1.582--1.756 & 1.207--1.310 \\
\bottomrule
\end{tabular}
\end{table}

\twocolumngrid

\end{document}